\documentclass[english,english,aps, prl, floatfix,twocolumn,reprint,showpacs]{revtex4}
\usepackage[T1]{fontenc}
\usepackage[latin9]{inputenc}
\usepackage{graphicx}
\usepackage{amssymb}

\makeatletter
\@ifundefined{textcolor}{}
{%
 \definecolor{BLACK}{gray}{0}
 \definecolor{WHITE}{gray}{1}
 \definecolor{RED}{rgb}{1,0,0}
 \definecolor{GREEN}{rgb}{0,1,0}
 \definecolor{BLUE}{rgb}{0,0,1}
 \definecolor{CYAN}{cmyk}{1,0,0,0}
 \definecolor{MAGENTA}{cmyk}{0,1,0,0}
 \definecolor{YELLOW}{cmyk}{0,0,1,0}
 }

\makeatother

\usepackage{babel}

\begin{document}

\title{NMR probing of the spin polarization of the $\nu=5/2$ quantum Hall
state}

\author{M. \surname{Stern}$^{1}$}
\email{mstern@weizmann.ac.il}
\author{B. A. \surname{Piot}$^{2}$}
\author{Y. \surname{Vardi}$^{1}$}
\author{V. \surname{Umansky}$^{1}$}
\author{P. \surname{Plochocka}$^{2}$}
\author{D. K. \surname{Maude}$^{2}$}
\author{I. \surname{Bar-Joseph}$^{1}$}

\affiliation{$^{1}$ Department of Condensed Matter Physics, The Weizmann Institute of Science, Rehovot, Israel}
\affiliation{$^{2}$ Laboratoire National des Champs Magn\'etiques Intenses, CNRS-UJF-UPS-INSA, 38042 Grenoble, France}

\date{\today }
\begin{abstract}
Resistively detected nuclear magnetic resonance is used to measure the Knight shift of the $^{75}$As nuclei and
determine the electron spin polarization of the fractional quantum Hall states of the second Landau level. We show that
the $5/2$ state is fully polarized within experimental error, thus confirming a fundamental assumption of the
Moore-Read theory. We measure the electron heating under radio frequency excitation, and show that we are able to
detect NMR at electron temperatures down to $30$~mK.
\end{abstract}

\pacs{73.43.Fj}

\maketitle The physics of interacting electrons at half integer filling factor has intrigued researchers over the past
two decades. It was found that the behavior at the first Landau level can be understood in terms of non-interacting
composite fermions (CF) at zero magnetic field \cite{Halperin}. Numerous experimental findings confirm this picture,
and establish the formation of a non-gaped state at $\nu=1/2$. The behavior in the second Landau level is, however,
fundamentally different. Early experiments, dating back to the late eighties, show the formation of a fractional
quantum Hall (FQH) state at $\nu=5/2$ \cite{Willett,Pan}. This observation has challenged the CF theory and triggered
an intense theoretical effort. Early on Moore and Read (MR) have suggested that a weak residual attractive interaction
in the second Landau level gives rise to pairing of the CFs, and to the formation of a gaped state \cite{Moore}. One of
the exciting aspects of MR theory is the predicted excitation spectrum of the $\nu=5/2$ state, which should consist of
quasiparticles that obey non-Abelian braiding statistics. It was shown that this property may turn the $5/2$ state into
a platform for quantum computing by means of topological manipulations \cite{DasSarma}.

The examination of MR theory has become the focus of intensive
experimental effort in recent years. A major support for its
validity has been provided by the measurement of the quasiparticle
charge, found to be e$/4$, in agreement with the prediction of the
MR theory \cite{Dolev,Radu}. However, this finding is also
consistent with other competing theories \cite{Halperin}, and
further experimental support is needed. The determination of the
electron spin polarization can provide this needed support. A
central assumption of the MR theory is that the electrons in the
second Landau level are fully-polarized, and can therefore form
pairs with p-type symmetry. Hence, confirming this point would
provide a strong experimental evidence for the validity of the MR
theory. Unfortunately, the experiments realized so far to probe
the polarization at $\nu=5/2$ have led to ambiguous results:
Tilted field experiments have shown that the $\nu=5/2$ gap
decreases with tilt angle~\cite{Eisenstein,Gervais}, which could
be interpreted in favor of a spin depolarized ground state.
However, this behavior may also originate from the destruction of
the $5/2$ state induced by the orbital coupling to the parallel
magnetic field~\cite{Morf,Rezayi,Peterson}. Optical measurements
also provided indications, which supported an unpolarized ground
state. Raman experiments have shown diminishing of the spin flip
mode, and were interpreted as an evidence for a partially
polarized second Landau level \cite{Rhone}. Recently, we reported
the results of photoluminescence (PL) measurements \cite{Stern}
that show a dip in the Zeeman splitting in the vicinity of
$\nu=5/2$, and can be interpreted as due to a spin unpolarized
ground state. However, a possible complication in the
interpretation of the optical data could arise from the role of
the positively charged valence band hole that is introduced by
laser illumination. The effect of this valence hole is not fully
understood and could, in principle, affect the observed results.
All these experimental findings are in clear contrast with the
results of numerical simulations which clearly show that the
ground state should be spin polarized \cite{Morf}. There are also
transport data which support a spin polarized state; \emph{e.g}
the observation of $\nu=5/2$ at high magnetic fields~\cite{Zhang}
and resistively detected NMR measurements performed at a
relatively high excitation power~\cite{Tiemann}. This controversy
calls for further experimental work using a different experimental
technique \cite{Jain}.

In this work we use resistively detected nuclear magnetic resonance (NMR) to measure the Knight shift of the $^{75}$As
nuclei and determine the electron spin polarization of the FQH states of the second Landau level. We monitor the
electron heating under radio frequency (RF) excitation, and show that we are able to detect an NMR signal at electron
temperatures down to 30 mK. We find that the FQH states in the second Landau level, and in particular the $5/2$ state,
are preserved under RF excitation. We show that the $5/2$ state is fully polarized, thus confirming a fundamental
assumption of the MR theory. Note that throughout this paper polarization refers to the polarization of the partially
occupied Landau level at the Fermi energy.

The NMR technique is a powerful tool to measure the electron spin
polarization. It is based on the coupling of the electron and
nuclei spins via the hyperfine interaction. In the presence of an
external magnetic field, the nuclei acquire an average
polarization $\left\langle I\right\rangle $ and create a local
magnetic field $B_{N}\varpropto\left\langle I\right\rangle $
(Overhauser effect). $B_{N}$ acts exclusively on the electronic
spin and has no influence on the orbital motion of the electrons,
so that the filling factor remains unchanged. The polarized
electrons also create a local magnetic field $B_{e}$ acting on the
nuclear spins. This field reduces the Larmor resonance frequency
of the nuclei by $K_{S}\varpropto n_{e}P$, an effect known as the
Knight shift, where $\emph{n}_{e}$ is the local electron density
and $\emph{P}$ is the electron spin polarization. Obtaining the
signal from a 2D electron gas is however an experimental
challenge, since the number of nuclei in a single quantum well is
much smaller than the total number of nuclei in the bulk of the
sample. Several techniques have been implemented to overcome this
difficulty; increasing the effective thickness by using multiple
quantum wells \cite{Melinte}, increasing the nuclei polarization
in the quantum well using optical pumping \cite{Barrett}, or using
a selective detecting scheme, which is specific to the electrons
in the well, known as resistively detected NMR
\cite{Dobers,Desrat}. This technique relies on the dependence of
the longitudinal resistance $R_{xx}$ on the Zeeman energy gap,
which can be written as $E_{Z}=g\mu_{B}\left(B+B_{N}\right)$. By
applying a radio frequency at the Larmor resonance frequency, it
is possible to depolarize the nuclear spins and reduce the
amplitude of $B_{N}$, and consequently decrease $E_{Z}$. This, in
turn, results in a slight change of $R_{xx}$, typically of the
order of a few percent.

\begin{figure}
\includegraphics[width=1\columnwidth]{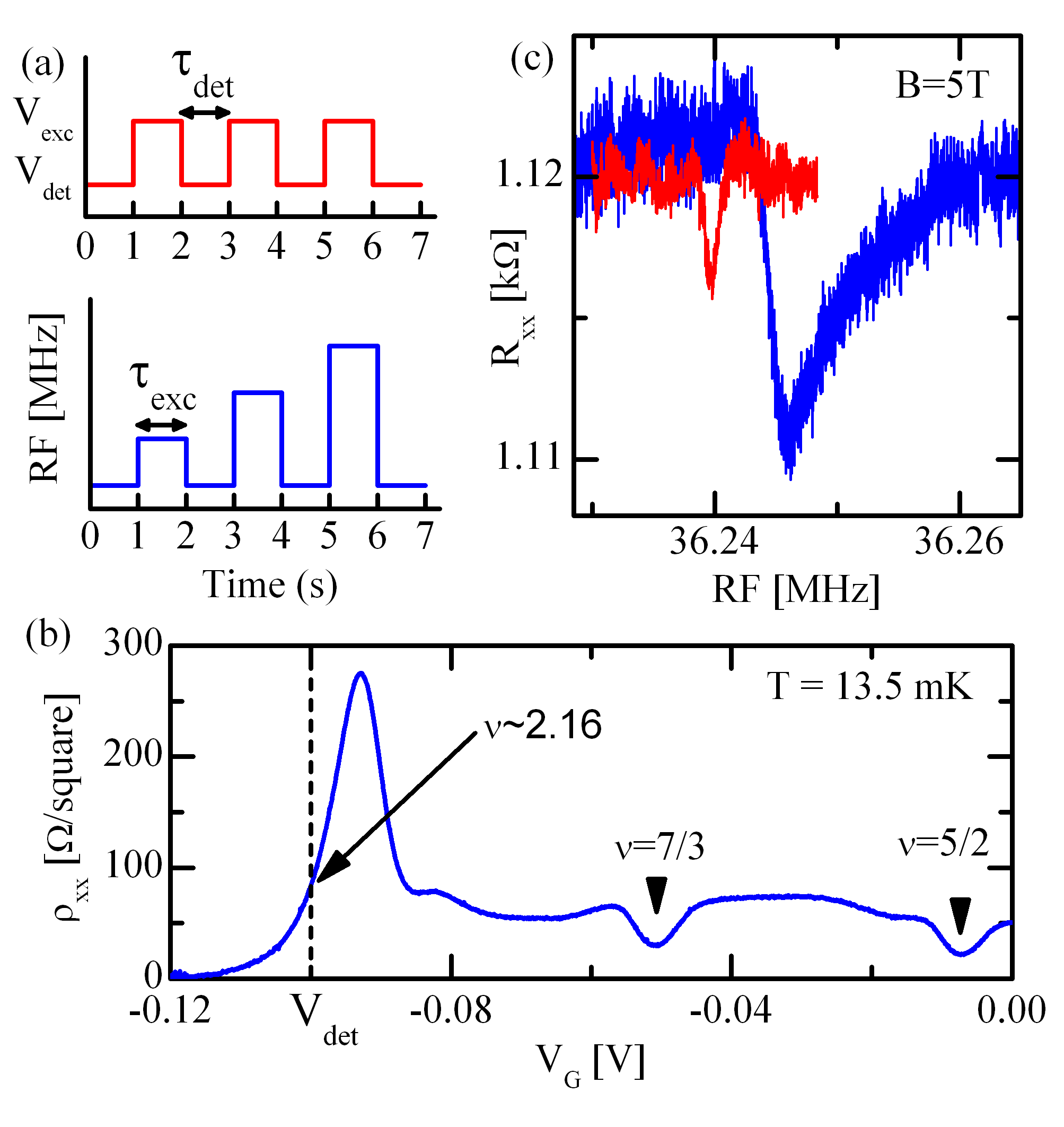}
\caption{(a) The NMR excitation and detection protocol as
described in the text.
%At initial time, the filling factor is tuned by applying a gate voltage \emph{$V_{exc}$}. After a short
%waiting time of 30 ms, the radio frequency is changed for a duration of $\tau_{exc}\sim1$s. If the applied frequency
%corresponds to the Larmor resonance frequency of the As nuclei, a depolarization of the nuclei occurs. To detect this
%depolarization, the gate voltage is turned back to the detection point $V_{det}$ and the change in the longitudinal
%resistance $R_{xx}$ is measured during $\tau_{det}\sim1$s, using a standard lock-in technique at frequency of $131$ Hz
%and current $I=30$ nA. The sequence is repeated many times while varying the radio frequency.
(b) $R_{xx}$ as a function of gate voltage showing the detection point (c) $R_{xx}$ versus applied radio frequency with
a power of -5 dBm; under excitation at $\nu=5/2$ and detection at $\nu=2.16$ (red line) and a scan at the detection
point (blue line). All measurements in this paper are performed at $B=5$~T.}\label{fig1}
\end{figure}

Attempts to directly implement this technique to measure the electron spin polarization at $\nu=5/2$ have failed; no
detectable change of $R_{xx}$ was measured throughout the quantum Hall plateau. Indeed, tilted field measurements show
that the longitudinal resistance at this filling factor has a very weak dependence on $E_{Z}$ \cite{Gervais}. One can,
however, measure the Knight shift at 5/2 by quickly switching to another filling factor at which $R_{xx}$ exhibits a
strong dependence on $E_{Z}$. This idea relies on the long relaxation time of the nuclei which decay back to their
original polarization on a timescale from a few tens of seconds to many minutes. It was implemented in
Ref.~\cite{Tiemann} by applying short RF pulses when the system is at $\nu=5/2$. We noticed, however, that applying
these RF pulses causes a non resonant transient response due to the heating of the electronic system and greatly
degrades the signal to noise ratio for the detection. For this reason, we have developed a technique where a low RF
power is \emph{continuously} applied. The improved signal to noise ratio allows us to work at much lower RF power, and
therefore lower electronic temperature than the pulsed NMR technique of Ref.\cite{Tiemann}. As will be shown this
reduction in the RF power is critical for the measurement.

The exact sequence used is shown in Fig.~\ref{fig1}(a). The detection of the signal is always performed under off
resonance RF at the detection filling factor $\nu=2.16$ (see Fig.~\ref{fig1}(b)). To excite the system, the filling
factor is set to the excitation point (\emph{e.g.} $\nu=5/2$). After a short waiting time ($\simeq 30$ms), the radio
frequency is abruptly changed during the excitation time $\tau_{exc}$, without changing the RF power. Then, the filling
factor and the radio frequency are returned to their original values and $R_{xx}$ is measured at the detection point
for a time interval of $\tau_{det}$. This sequence is repeated many times while varying the excitation radio frequency
so as to sweep through the resonance. A few important considerations should be mentioned here:

(i) The quick change of the filling factor between the excitation
and detection point can only be achieved by applying a gate
voltage. The sample we used is the \textit{same} we used in our
previous work, in which we measured the PL \cite{Stern}, and has a
4 nm PdAu gate.

(ii) The detection point should be close enough to the filling factor under consideration, to avoid irreversible
deterioration of the quality of the FQH states that occurs under strong gating. We have selected to work at the high
end of the $\nu=2$ plateau, at $\nu\simeq2.16$. In this region $R_{xx}$ exhibits a significant dependence on the Zeeman
energy and the NMR signal can be easily observed in a standard continuous wave experiment.

(iii) The RF power couples also to the electronic system and heats
it up. Thus, one should conduct in parallel a measurement of the electron
temperature and confirm that the quantum Hall state is not destroyed.

(iv) The long time for full thermalization of the nuclei, which at
5 T can be days, may inhibit performing sequential frequency
scans. Thus, one should provide means for fast nuclear
repolarization between frequency scans. This is effectively done
by ramping the magnetic field to $\sim11$ T, at the end of each
scan.

(v) We found that the ramping of the field can induce a change of
the value of the Larmor frequency of a $\simeq1$~kHz. Hence, it
is essential to \emph{always} have a  point to which each
scan can be compared. This is done by performing a  measurement
at the detection point at the end of each scan. This procedure sets
a limit on the precision of measuring the Knight shift to be 300~Hz.

Figure~\ref{fig1}(c) demonstrates the implementation of this
technique. The red curve shows the $R_{xx}$ signal under
excitation at 5/2 and detection at 2.16. It is seen that a well
resolved dip with an amplitude of 4 $\Omega$ appears at 36.24 MHz.
The blue curve is a scan obtained by performing a regular
resistively detected NMR measurement at the detection point
($\nu=2.16$). The fact that the $5/2$ dip appears at lower
frequency than the signal at $\nu=2.16$ is significant and a
qualitative conclusion can already be drawn. Since the electron
polarization can only reduce the Larmor frequency ($K_{S}\leq0$ )\
it follows that the $\nu=5/2$ FQH state cannot be fully
depolarized, in contrast with the results of the PL measurements.
Note that all our measurements are shifted compared to the
accepted value of the gyromagnetic ratio for $^{75}$As in bulk
GaAs (7.29) due to the remanent field (-27 mT) of the magnet which
reduces the effective magnetic field to $4.973$~T.

\begin{figure}
\includegraphics[width=1\columnwidth]{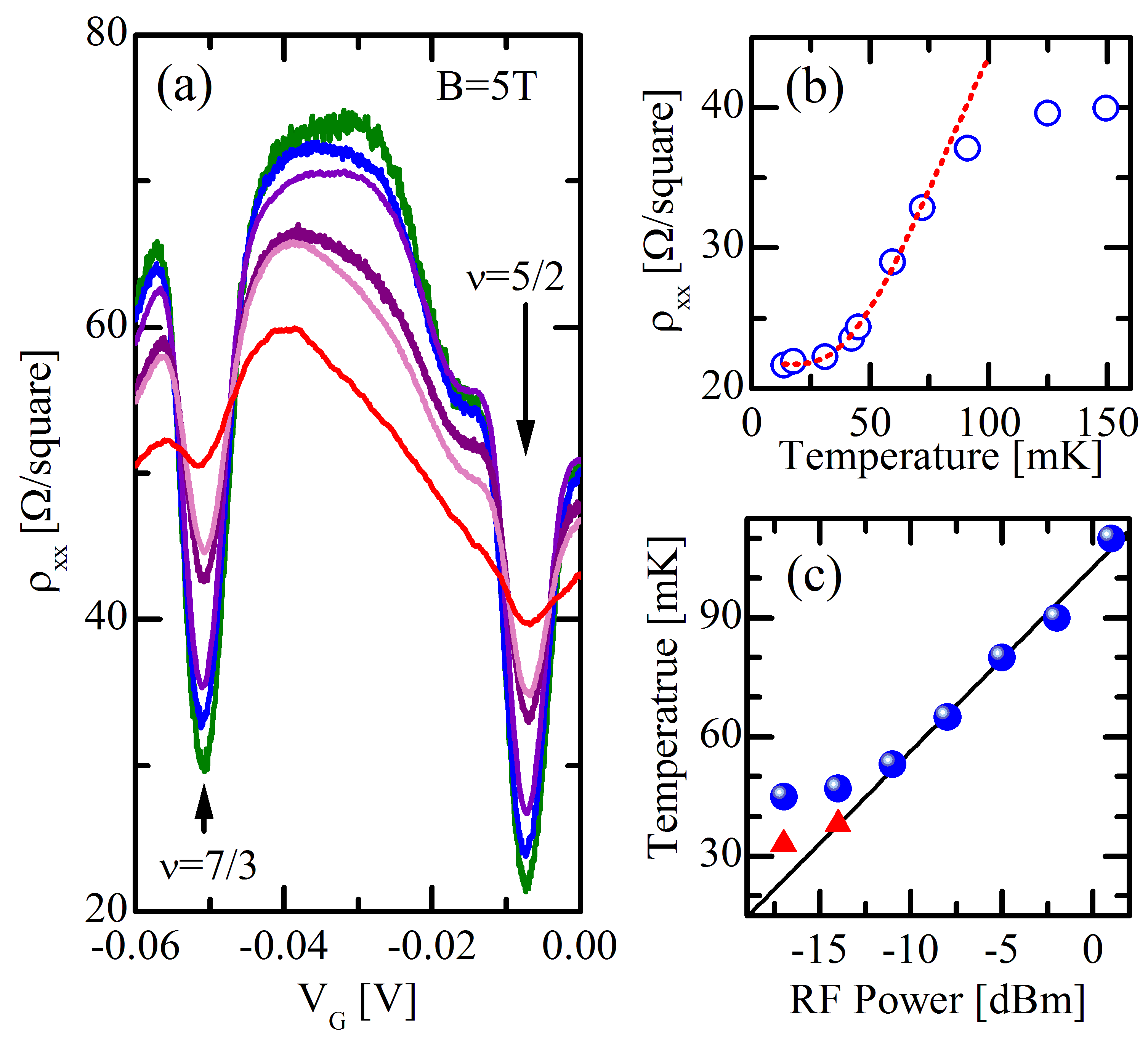}
\caption{(a) Resistivity $\rho_{xx}$ at $B=5$~T versus gate
voltage $V_{G}$ for various RF powers: +1 dBm, -2 dBm, -8 dBm,
-11dBm and -14 dBm. The black curve is $\rho_{xx}$ at $T=13.5$~mK
without RF power. (b) $\rho_{xx}$ at filling factor $\nu=5/2$
versus the electron temperature. The dotted red line is a fit to
$\rho_{xx}=\rho_{0}+\rho_{1}\exp\left[-\Delta/2T\right]$ giving
$\Delta\simeq300$ mK. (c) Electron temperature versus RF power.
Below -14dBm, the temperature is limited by the current $I=30$~nA.
The red triangles correspond to $I=5$~nA. }\label{fig2}
\end{figure}

Before presenting a quantitative analysis of the shift and its implications, one should verify that the observed
polarization is not due to heating of the electrons by the RF power and that the FQH state still exists.
Figure~\ref{fig2}(a) shows the longitudinal resistance of the sample as the RF power is reduced, from 1 dBm down to -14
dBm. It is seen that the resistance at $7/3$ and $5/2$ increases as the RF power is increased until the minima become
barely visible at 1 dBm . To obtain a calibration of the electron temperature we measured the dependence of $\rho_{xx}$
on the bath temperature at several filling factors without RF excitation. We show in Fig.~\ref{fig2}(b) the behavior at
$\nu=5/2$. This allows to obtain a calibration curve of the electron temperature as a function of the RF power
(Fig.\ref{fig2}(c)). It is seen that at 0 dBm the electrons are heated up to 100 mK. At the lowest RF powers the curve
deviates from a linear dependence (solid line). At this regime the electron temperature is limited by the applied
current (30 nA), and reducing it to 5 nA allows us to bring the electron temperature down to 33 mK. The residual
heating by the RF power is demonstrated in Fig.~\ref{fig2}(a); the lowest curve is taken without RF power and it is
seen that its resistance is only slightly below that measured at -14 dBm.

To evaluate the effect of the residual heating of the electrons
one needs to estimate the 5/2 gap. The dotted line in
Fig.\ref{fig2}(b) is a fit to
$\rho_{xx}=\rho_{0}+\rho_{1}\exp\left[-\Delta/2T\right]$. Here
$\rho_{0}=22$ $\Omega$\ is the lowest value of $\rho_{xx}$ at zero
temperature which is different than zero for this sample,
$\rho_{1}=120$ $\Omega$, and $\Delta\simeq300$ mK. We can
therefore conclude that the region below $-5$ dBm corresponds to
the activation region, and by reducing the RF power we can
determine the behavior of the spin polarization down to
$T/\Delta\simeq0.1$.

\begin{figure}
\includegraphics[width=1\columnwidth]{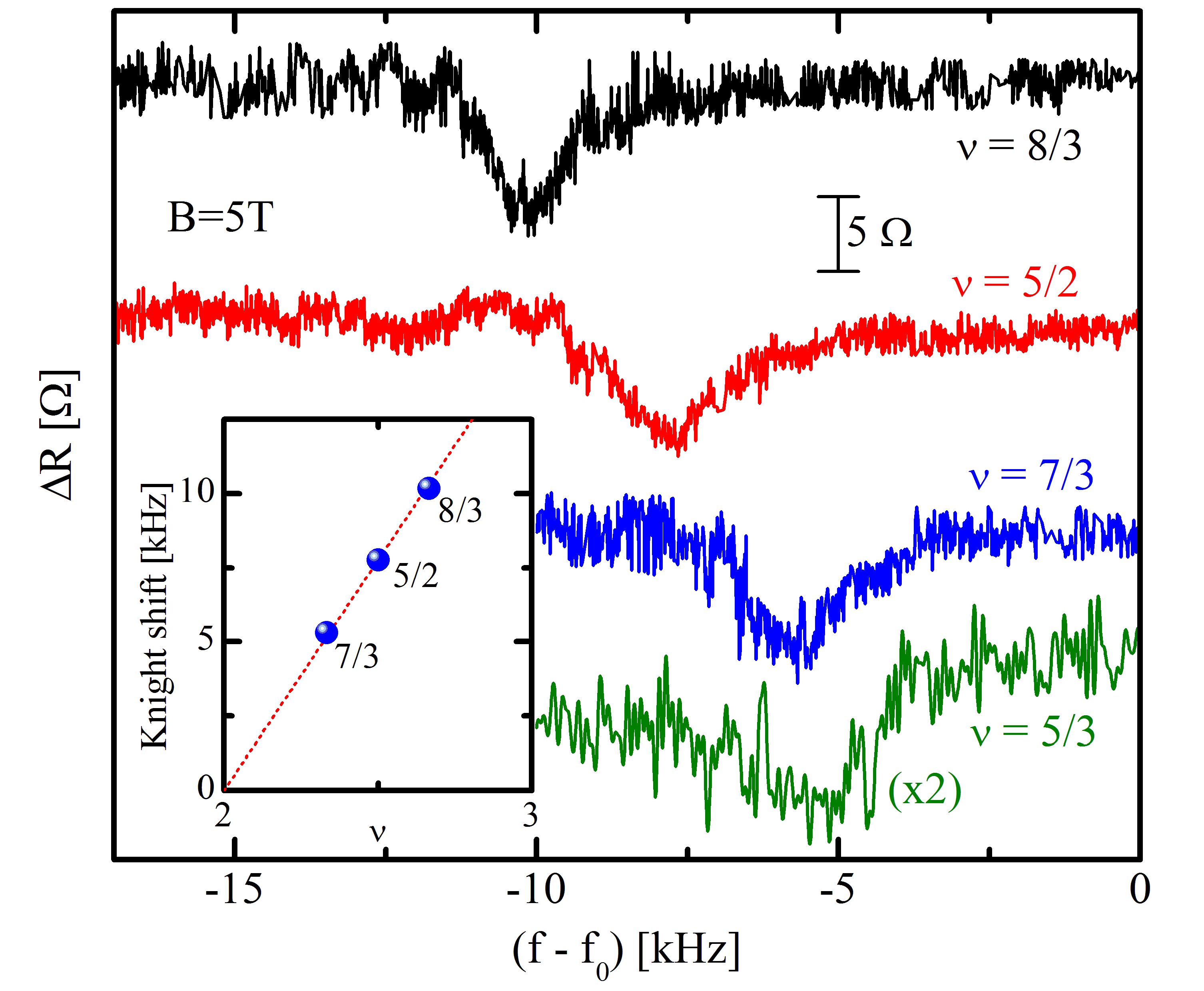}
\caption{Change of the longitudinal resistance $\Delta R$ as a
function of radio frequency ($f-f_0$) where $f_0$ is the resonance
frequency at $\nu=2$. The RF power is here -11 dBm. Inset: Knight
shift as a function of filling factor. The red dotted line
corresponds to full electron polarization according to the shift
obtained at $\nu=5/3$.}\label{fig3}
\end{figure}

Figure~\ref{fig3} shows the change in resistance, $\Delta R$, as a
function of radio frequency $f-f_0$ at four filling factors, 5/3,
7/3, 5/2, and 8/3, where $f_0$ is the resonance frequency at
$\nu=2$. Since the $\nu=2$ state is clearly depolarized, it will
have a resonance frequency $f_0$ \emph{i.e} the bare (unshifted)
Larmor frequency. On the other hand, the fractional quantum Hall
state at $\nu=5/3$ is the electron-hole symmetric state of
$\nu=1/3$ and is known to be fully spin polarized \cite{Clark}. As
we were unable to measure a convincing signal at $\nu=2$ with our
technique at low power a lengthy series of calibration
measurements were performed at higher RF power using a pulsed NMR method similar to
Ref.\cite{Tiemann}. During the calibration procedure we
simultaneously measured the signal at $\nu=5/3$ to obtain a
calibration of the maximum possible Knight shift corresponding to
a full degree of polarization. As $\nu=5/3$ is also detected in
the low power measurements this provides a direct calibration for
$f0$ in our measurements. Additionally, this calibration was
confirmed by measurements at $\nu=2/3$ which can serve as a
reference for both fully polarized and unpolarized states due to
the formation of domains~\cite{SternPRB}.

%The resonances at these two filling factors have been used in these measurements as a reference for zero and full
%polarization and were precisely measured at high RF power.

From our measurements we obtain a different $K_{S}$ equal to $5650\pm75$ Hz, $7750\pm175$ Hz and $10175\pm150$ Hz for
$\nu=7/3$, 5/2 and 8/3, respectively. In the inset of Fig.~\ref{fig3} we plot $K_{S}$ as a function $\nu$; remarkably,
we find that they reside on a straight line, implying the same degree of spin polarization. The red line corresponds to
the extrapolated value of the Knight shift for a fully polarized state according to its value at $\nu=5/3$.  We can,
thus, conclude that the electrons at the second Landau level are fully polarized at the three filling factors 7/3, 5/2
and 8/3.

\begin{figure}
\includegraphics[width=1\columnwidth]{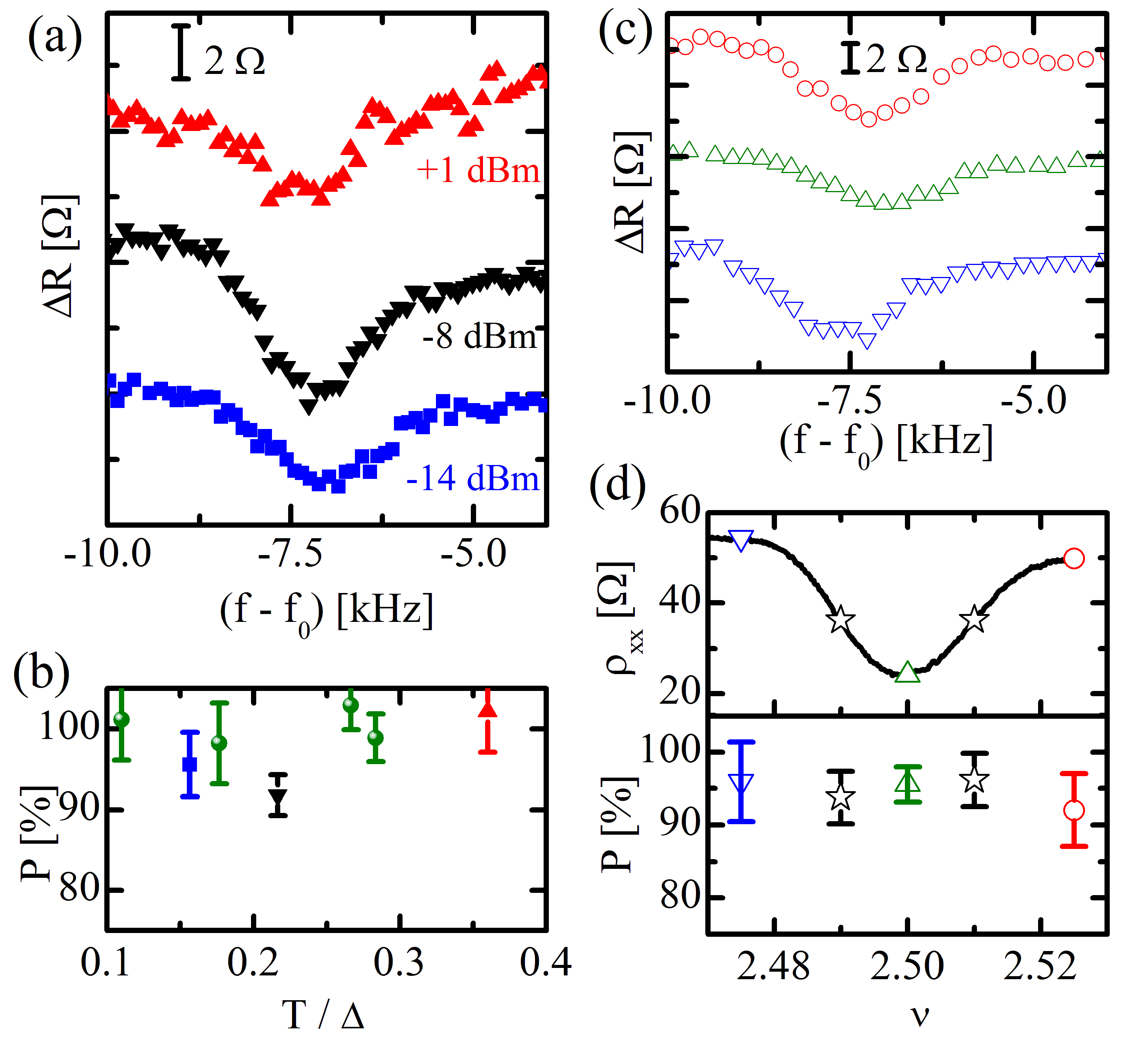}
\caption{(a) NMR signal at $\nu=5/2$ for different RF powers. (b)
Electron spin polarization at $\nu=5/2$ as a function of
$T/\Delta$. To further reduce the heat load on the sample at the
lowest RF powers, the current is reduced to $I=1$~nA during the
excitation phase. NMR signal (c) and the resistivity together with
the electron spin polarization (d) in the vicinity of $\nu=5/2$ at
RF power -14 dBm. Note the correspondence of the symbols in (a-b)
and in (c-d).} \label{fig4}
\end{figure}

The full polarization at 5/2 is robust and persists throughout the temperature range that we have studied, and
throughout the Hall plateau, as demonstrated in Fig.\ref{fig4}. Figure~\ref{fig4}(a) shows $\Delta R$ as a function of
frequency for three RF powers, 1 dBm, -8 dBm and -14 dBm, which correspond to electron temperatures of 110, \ 65, and
35 mK, respectively. It is readily seen that the resonance frequency does not change and the spin polarization remains
constant at $0.1<T/\Delta<0.35$ (Fig.\ref{fig4}(b). Figure~\ref{fig4}(c-d) show a measurement at -14 dBm in the
vicinity of $\nu=5/2$; no depolarization is observed within the $5\%$ precision of our measurement.

Our results confirm the assumption of the MR theory that the electrons are fully polarized in the ground state.
However, they seem to be in stark contradiction with the results of the photoluminescence measurements \cite{Stern},
which were conducted on the same sample. We note that the two measurements probe different aspects of the FQH state.
The NMR measurement probes the spin polarization through its effect on the longitudinal resistance. Since the change in
$R_{xx}$ are related to tunneling through saddle points between edges at the two sides of the sample
\cite{d'Ambrumenil,Nuebler} - the effect is global and is related to the average spin polarization of the whole sample.
The PL measurement, on the other hand, is local in nature, and samples the immediate neighborhood of the valence hole.
The results of the PL measurement imply that this local environment is depolarized. Recently, it was suggested that at
$\nu=5/2$ Skyrmions, which consist of two quasiparticles with opposite spin, tend to form in local minima of the
disordered potential (formed by well width disorder or remote impurities)~\cite{Wojs}. One can therefore imagine two
possible scenarios that might explain the origin of the observed depolarization. The first is that the holes are
attracted to the same local minima at which Skyrmions are formed. The second is that the potential of the valence band
hole, which induces a potential minima, could help forming a Skyrmion, essentially playing the same role as the
disorder in Ref.\cite{Wojs}. In both cases the immediate environment of the valence hole is unpolarized. We note that
this is a unique property of the 5/2 state, and in this sense the optical measurement probes the Skyrmion formation.
\begin{acknowledgments}
We thank A. Stern, A. Wojs and S. Simon for fruitful discussions,
and L. Sepunaru for his help in the first characterization of the
samples. This research was supported by the Israeli Science Foundation. \end{acknowledgments}


\begin{thebibliography}{24}
\bibitem{Halperin} B. I. Halperin, P.A. Lee and N. Read, Phys. Rev.
B, \textbf{47}, 7312 (1993).
\bibitem{Willett} R. Willett \textit{et al.}, Phys. Rev. Lett., \textbf{59},
1776 (1987).
\bibitem{Pan} W. Pan \emph{et al.} Phys. Rev. Lett. \textbf{83},
3530 (1999).
\bibitem{Moore} G. Moore, N. Read, Nucl. Phys. B, \textbf{360}, 362
(1991).
\bibitem{DasSarma} S. Das Sarma, M. Freedman \& C. Nayak, Phys. Rev.
Lett., \textbf{94}, 166802 (2005).
\bibitem{Dolev} M. Dolev \textit{et al.}, Nature, \textbf{452} 829
(2008).
\bibitem{Radu} I. P. Radu \textit{et al.}, Science, \textbf{320},
899 (2008).
\bibitem{Eisenstein} J.P. Eisenstein \textit{et al.}, Phys. Rev.
Lett., \textbf{61}, 997 (1988).
\bibitem{Gervais} C. Dean \textit{et al.}, Phys. Rev. Lett., \textbf{101},
186806 (2008).
\bibitem{Morf} R. H. Morf, Phys. Rev. Lett., \textbf{80}, 1505 (1998).
\bibitem{Rezayi} E. H. Rezayi and F. D. M. Haldane, Phys. Rev. Lett. \textbf{84},
4685 (2000).
\bibitem{Peterson} M. R. Peterson, T. Jolicoeur, S. DasSarma, Phys. Rev. Lett.
\textbf{101}, 016807 (2008).
\bibitem{Rhone} T. D. Rhone \emph{et al.}, Phys. Rev. Lett., 106,
196805 (2011).
\bibitem{Stern} M. Stern \emph{et al.}, Phys. Rev. Lett., \textbf{105},
096801 (2010).
\bibitem{Zhang} C. Zhang et al. Phys. Rev. Lett. \textbf{104}, 166801
(2010).
\bibitem{Tiemann} L. Tiemann, G. Gamez, N. Kumada, and K. Muraki,
Meeting abstracts of the Physical Society of Japan, \textbf{65}, 649
(2010).
\bibitem{Jain} J.K. Jain, Physics, \textbf{3}, 71 (2010).
\bibitem{Melinte} S. Melinte \emph{et al.}, Phys. Rev. Lett\emph{.},
\textbf{84}, 354 (2000).
\bibitem{Barrett} S.E. Barrett, R. Tycko, L.N. Pfeiffer and K.W.
West, Phys. Rev. Lett\emph{.},\textbf{72}, 1368 (1994).
\bibitem{Dobers} M. Dobers, K. v. Klitzing, J. Schneider, G. Weimann and K. Ploog, Phys. Rev. Lett., \textbf{61},
1650 (1988).
\bibitem{Desrat} W. Desrat \emph{et al.}, Phys. Rev. Lett., \textbf{88},
256807 (2002).
\bibitem{Clark} R. G. Clark \emph{et al.}, Phys. Rev. Lett., \textbf{62}, 1536 (1989).
\bibitem{SternPRB} O. Stern \emph{et al.} Phys. Rev. B \textbf{70}, 075318 (2004).
\bibitem{Nuebler} J. Nuebler \emph{et al.}, Phys. Rev. B, \textbf{81},035316
(2010).
\bibitem{d'Ambrumenil} N. d'Ambrumenil, B. I. Halperin and R. H.
Morf, Phys. Rev. Lett., \textbf{106}, 126804 (2011).
\bibitem{Wojs} A. Wojs, G. Moller, S. H. Simon, and N. R. Cooper,
Phys. Rev. Lett., \textbf{104}, 086801 (2010).


\end{thebibliography}
\end{document}